**Approximate solutions of the Schrödinger equation with Hulthén-Hellmann Potentials for a Quarkonium system**

I. O. Akpan, E. P. Inyang, E. P. Inyang and E. S. William

*Theoretical Physics Group, Department of Physics, University of Calabar, P.M.B 1115, Calabar, Nigeria*

*Corresponding author email: etidophysics@gmail.com*

**Abstract**

Hulthén plus Hellmann potentials are adopted as the quark-antiquark interaction potential for studying the mass spectra of heavy mesons. We solved the radial Schrödinger equation analytically using the Nikiforov-Uvarov method. The energy eigenvalues and corresponding wave function in terms of Laguerre polynomials were obtained. The present results are applied for calculating the mass of heavy mesons such as charmonium $c\bar{c}$ and bottomonium $b\bar{b}$. Four special cases were considered when some of the potential parameters were set to zero, resulting into Hellmann potential, Yukawa potential, Coulomb potential, and Hulthén potential, respectively. The present potential provides satisfying results in comparison with experimental data and the work of other researchers.

### 1. Introduction

The study of the fundamental or constituent blocks of matter has been for long time a fascinating field in physics. In the nineteenth century, the atom was considered to be the fundamental particles from which all matters were composed. This idea was used to explain the basic structure of all elements [1].

The problem of what were considered to be fundamental particles was resolved by the quarks. Because of the heavy masses of the constituent quarks, a good description of many features of these systems can be obtained using non-relativistic models that is the quark-antiquark strong interaction is described by a phenomenological potential [2]. Heavy quarkonium systems have turned out to provide extremely useful probes for the deconfined state of matter because the force between a heavy quark and anti-quark is weakened due to the presence of gluons which lead to the dissociation of quarkonium bound states [3]. The quarkonia with heavy quark and antiquark and their interaction are well described by the Schrödinger equation (SE) [4]. The solution of the spectral problem for the SE with spherically symmetric potentials is of major concern in describing the spectra of quarkonia [5]. Potential models offer a rather good description of the mass spectra of systems such as a bottomonium, and charmonium [6]. In simulating the interaction potentials for these systems, confining-type potentials are generally used. The holding potential is the so called Cornell potential with two terms one of which is responsible for the Coulomb interaction of the quarks and the other corresponds to a confining term [7].

The solutions to the SE can be established only if we know the confining potential for a particular physical system. Till now, there are only a few confining potentials, like the harmonic oscillator and the hydrogen atom, for which solutions to the SE are found exactly [8].

The Hulthén potential takes the form [9]

$$V(r) = -\frac{A_0 e^{-\alpha r}}{1 - e^{-\alpha r}} \tag{1}$$

where $\alpha$ is the screening parameter and $A_0$ is the potential strength which is sometimes identified with the atomic number when the potential is used for atomic phenomena [10]. It is a short-range potential which is applied in many branches of physics, such as nuclear and particle physics, atomic physics, solid state physics, and chemical physics [11,12].

The Hellmann potential which is a superposition of an attraction Coulomb potential and a Yukawa potential can be expressed as [13].

$$V(r) = -\frac{A_1}{r} + \frac{A_2 e^{-\alpha r}}{r} \tag{2}$$

where the parameters $A_1$ and $A_2$ denote the strength of Coulomb and Yukawa potentials respectively, $\alpha$ denotes the screening parameter, and $r$ is the distance between two particles. These potentials have been used to study bound state problems from many researchers [14-20]. Recently, Inyang et al.[21] obtained the Klein-Gordon equation solutions for the Yukawa potential using the Nikiforov-Uvarov method. The energy eigenvalues were obtained both in a relativistic and non-relativistic regime. They applied the results to calculate heavy-meson masses of charmonium $c\bar{c}$ and bottomonium $b\bar{b}$. Apart from that many researchers have provided approximate solutions to SE using different methods with Cornell potential. For instance, Vega and Flores [22] obtained the approximate solutions of the Schrödinger equation with the Cornell potential using variational method and super symmetric quantum mechanics (SUSYQM). Abu-Shady et al.[23] studied the N-dimensional radial Schrödinger equation using the analytical exact iteration method (AEIM), in which the Cornell potential is generalized to finite temperature and chemical potential. In addition, Ciftci and Kisoglu [24], solved non-relativistic arbitrary $l$-states of quarkonium through asymptotic iteration method (AIM). An analytic solution of the N-dimensional radial Schrödinger equation with the mixture of vector and scalar potentials via the Laplace transformation method (LTM) were studied by [25]. Al-Jamel and Widyan [26] studied heavy quarkonium mass spectra in a Coulomb field plus quadratic potential using NU method. Ibekwe et al.[27] solved the radial SE with an exponential, generalized, harmonic Cornell potential using the series expansion method. Their results were used to calculate the mass spectra of heavy-mesons. Al-Oun et al.[28] examine heavy quarkonia characteristics properties in the general framework of non-relativistic potential model consisting of a Coulomb plus quadratic potential. Chouikh et al.[29] proposed an approach to achieve quantum computation with atomics qubits in a cavity QED. Recently, researchers have shown great interest in the combination of two or more potentials in both the relativistic and non relativistic approach. The fundamental nature of combining two or more physical potential models is to have a wider range of application [30]. For example, Cornell potential which is the combination of Coulomb potential with linear terms is used in studying the mass spectra for coupled states and for the electromagnetic characteristics of meson [31]. For instance, William et al.[32] obtained bound state solutions of the radial Schrödinger equation by the combination of Hulthén and Hellmann potential within the framework of Nikiforov-Uvarov (NU) method. Also, Edet et al.[33] obtained an approximate solution of the SE for the modified Kratzer potential plus screened Coulomb potential model using the Nikiforov-Uvarov (NU) method. In this present work, we aim to study the SE with the combination of Hulthén and Hellmann potential analytically by using NU method and apply the results to calculate the mass spectra of heavy quarkonium particles such as bottomonium and charmonium, in which the quarks are considered as spinless particles for easiness, which have not been considered before using this potentials to the best of our knowledge. The adopted potential is of the form [32]

$$V(r) = -\frac{A_0 e^{-\alpha r}}{1 - e^{-\alpha r}} - \frac{A_1}{r} + \frac{A_2 e^{-\alpha r}}{r} \tag{3}$$

where $A_0, A_1$ and $A_2$ are potential strength parameters and $\alpha$ is the screening parameter. In other to make Eq.(3) temperature dependent, the screening parameter is replaced with Debye mass $(m_D(T))$ which is temperature dependent and vanishes at T→ 0 and we have,

$$V(r,T) = -\frac{A_0 e^{-m_D(T)r}}{1 - e^{-m_D(T)r}} - \frac{A_1}{r} + \frac{A_2 e^{-m_D(T)r}}{r} \tag{4}$$

We carry out series expansion of the exponential terms in Eq.(4) up to order three, in order to model the potential to interact in the quark-antiquark system and this yields,

$$\frac{e^{-m_D(T)r}}{r} = \frac{1}{r} - m_D(T) + \frac{m_D^2(T)r}{2} - \frac{m_D^3(T)r^2}{6} + \ldots \tag{5}$$

$$\frac{e^{-m_D(T)r}}{1 - e^{-m_D(T)r}} = \frac{1}{m_D(T)r} - \frac{1}{2} + \frac{m_D(T)r}{12} + \ldots \tag{6}$$

We substitute Eqs.(5) and (6) into Eq.(4) and obtain

$$V(r,T) = -\frac{\beta_0}{r} + \beta_1 r - \beta_2 r^2 + \beta_3 \tag{7}$$

where

$$\left. \begin{array}{l} -\beta_0 = A_2 - A_1 - \dfrac{A_0}{m_D(T)}, \quad \beta_1 = \dfrac{A_2 m_D^2(T)}{2} - \dfrac{A_0 m_D(T)}{12} \\ \\ \beta_2 = \dfrac{A_2 m_D^3(T)}{6}, \quad \beta_3 = \dfrac{A_0}{2} - A_2 m_D(T) \end{array} \right\} \tag{8}$$

The first term in Eq. (7) is the Coulomb potential that describes the short distance between quarks, while the second term is a linear term for confinement feature.

## 2. Approximate solutions of the Schrödinger equation with Hulthén plus Hellmann potential

The Schrödinger equation (SE) for two particles interacting via potential $V(r)$ in three dimensional space, is given by [34]

$$\frac{d^2 R(r)}{dr^2} + \left[ \frac{2\mu}{\hbar^2}(E_{nl} - V(r)) - \frac{l(l+1)}{r^2} \right] R(r) = 0 \tag{9}$$

where $l, \mu, r$ and $\hbar$ are the angular momentum quantum number, the reduced mass for the quarkonium particle, inter-particle distance and reduced plank constant respectively. We substitute Eq.(7) into Eq.(9) and obtain

$$\frac{d^2 R(r)}{dr^2} + \left[ \frac{2\mu E_{nl}}{\hbar^2} + \frac{2\mu \beta_0}{\hbar^2 r} - \frac{2\mu \beta_1 r}{\hbar^2} + \frac{2\mu \beta_2 r^2}{\hbar^2} - \frac{2\mu \beta_3}{\hbar^2} - \frac{l(l+1)}{r^2} \right] R(r) = 0 \tag{10}$$

Let,

$$\zeta = \frac{2\mu}{\hbar^2}(E_{nl} - \beta_3), \quad \alpha_0 = \frac{2\mu\beta_0}{\hbar^2}$$
$$\alpha_1 = \frac{2\mu\beta_1}{\hbar^2}, \quad \alpha_2 = \frac{2\mu\beta_2}{\hbar^2}, \quad \gamma = l(l+1) \quad (11)$$

Substituting Eq.(11) into Eq.(10), we have

$$\frac{d^2 R(r)}{dr^2} + \left[\zeta + \frac{\alpha_0}{r} - \alpha_1 r + \alpha_2 r^2 - \frac{\gamma}{r^2}\right] R(r) = 0 \quad (12)$$

Transforming the coordinate of Eq.(12) we set

$$x = \frac{1}{r} \quad (13)$$

Differentiating Eq.(13) and simplifying we have

$$\frac{d^2 R}{dr^2} = \frac{2}{r^3} \frac{dR}{dx} + \frac{1}{r^4} \frac{d^2 R}{dx^2} \quad (14)$$

Substituting Eqs.(13) and (14) into Eq.(12) we have

$$\frac{d^2 R(x)}{dx^2} + \frac{2}{x} \frac{dR}{dx} + \frac{1}{x^4}\left[\zeta + \alpha_0 x - \frac{\alpha_1}{x} + \frac{\alpha_2}{x^2} - \gamma x^2\right] R(x) = 0 \quad (15)$$

Next, we propose the following approximation scheme on the term $\frac{\alpha_1}{x}$ and $\frac{\alpha_2}{x^2}$.

Let us assume that there is a characteristic radius $r_0$ of the meson. Then the scheme is based on the expansion of $\frac{\alpha_1}{x}$ and $\frac{\alpha_2}{x^2}$ in a power series around $r_0$; i.e. around $\delta \equiv \frac{1}{r_0}$, in the x-space up to the second order. This is similar to Pekeris approximation, which helps to deform the centrifugal term such that the modified potential can be solved by NU method [35].

Setting $y = x - \delta$ and around $y = 0$ it can be expanded into a series of powers as;

$$\frac{\alpha_1}{x} = \frac{\alpha_1}{y + \delta} = \frac{\alpha_1}{\delta\left(1 + \frac{y}{\delta}\right)} = \frac{\alpha_1}{\delta}\left(1 + \frac{y}{\delta}\right)^{-1} \quad (16)$$

which yields

$$\frac{\alpha_1}{x} = \alpha_1\left(\frac{3}{\delta} - \frac{3x}{\delta^2} + \frac{x^2}{\delta^3}\right) \quad (17)$$

Similarly,

$$\frac{\alpha_2}{x^2} = \alpha_2\left(\frac{6}{\delta^2} - \frac{8x}{\delta^3} + \frac{3x^2}{\delta^4}\right) \quad (18)$$

By substituting Eqs.(18) and (17) into Eq.(15), we obtain

$$\frac{d^2 R(x)}{dx^2} + \frac{2x}{x^2}\frac{dR(x)}{dx} + \frac{1}{x^4}\left[-\varepsilon + \alpha x - \beta x^2\right]R(x) = 0 \tag{19}$$

where

$$-\varepsilon = \left(\zeta + \frac{6\alpha_2}{\delta^2} - \frac{3\alpha_1}{\delta}\right), \quad \alpha = \left(\frac{3\alpha_1}{\delta^2} + \alpha_0 - \frac{8\alpha_2}{\delta^3}\right), \quad \beta = \left(\gamma + \frac{\alpha_1}{\delta^3} - \frac{3\alpha_2}{\delta^4}\right) \tag{20}$$

Comparing Eq.(19) and Eq.(A1) we obtain

$$\left.\begin{array}{l}\tilde{\tau}(x) = 2x, \quad \sigma(x) = x^2 \\ \tilde{\sigma}(x) = -\varepsilon + \alpha x - \beta x^2 \\ \sigma'(x) = 2x, \quad \sigma''(x) = 2\end{array}\right\} \tag{21}$$

We substitute Eq.(21) into Eq.(A9) and obtain

$$\pi(x) = \pm\sqrt{\varepsilon - \alpha x + (\beta + k)x^2} \tag{22}$$

To determine $k$, we take the discriminant of the function under the square root, which yields

$$k = \frac{\alpha^2 - 4\beta\varepsilon}{4\varepsilon} \tag{23}$$

We substitute Eq.(23) into Eq.(22) and have

$$\pi(x) = \pm\left(\frac{\alpha x}{2\sqrt{\varepsilon}} - \frac{\varepsilon}{\sqrt{\varepsilon}}\right) \tag{24}$$

For a physically acceptable solution, we take the negative part of Eq.(24) which is required for bound state problems and differentiate, this yields

$$\pi'_-(x) = -\frac{\alpha}{2\sqrt{\varepsilon}} \tag{25}$$

Substituting Eqs.(21) and (25) into Eq.(A7) we have

$$\tau(x) = 2x - \frac{\alpha x}{\sqrt{\varepsilon}} + \frac{2\varepsilon}{\sqrt{\varepsilon}} \tag{26}$$

Differentiating Eq.(26) we have

$$\tau'(x) = 2 - \frac{\alpha}{\sqrt{\varepsilon}} \tag{27}$$

By using Eq.(A10), we obtain

$$\lambda = \frac{\alpha^2 - 4\beta\varepsilon}{4\varepsilon} - \frac{\alpha}{2\sqrt{\varepsilon}} \tag{28}$$

And using Eq.(A11), we obtain

$$\lambda_n = \frac{n\alpha}{\sqrt{\varepsilon}} - n^2 - n \tag{29}$$

Equating Eqs.(28) and (29), the energy eigenvalues of Eq.(10) is given

$$E_{nl} = A_0 \left( \frac{1}{2} - \frac{m_D(T)}{4\delta} \right) + A_2 m_D(T) \left( \frac{3m_D(T)}{2\delta} - m_D^2(T) - 1 \right)$$

$$- \frac{\hbar^2}{8\mu} \left[ \frac{\frac{2\mu}{\hbar^2}\left(A_2 - A_1 + \frac{A_0}{m_D(T)}\right) + \frac{\mu m_D(T)}{\hbar^2 \delta^2}\left(3A_2 m_D(T) - \frac{A_0}{2}\right) - \frac{8\mu A_2 m_D^3(T)}{3\hbar^2 \delta^3}}{n + \frac{1}{2} + \sqrt{\left(l + \frac{1}{2}\right)^2 + \frac{\mu A_2 m_D^2(T)}{\hbar^2 \delta^3}\left(1 - \frac{m_D(T)}{\delta}\right) - \frac{\mu A_0 m_D(T)}{6\hbar^2 \delta^3}}} \right]^2 \tag{30}$$

2.1 Special cases

In this subsection, we obtain the special case by setting some parameters to zero.

1. When we set $A_0 = A_1 = 0$, we obtain the energy eigenvalues for Yukawa potential

$$E_{nl} = A_2 m_D(T)\left(\frac{3m_D(T)}{2\delta} - m_D^2(T) - 1\right) - \frac{\hbar^2}{8\mu}\left[\frac{\frac{-2\mu A_2}{\hbar^2} + \frac{3\mu A_2 m_D^2(T)}{\hbar^2 \delta^2} - \frac{8\mu A_2 m_D^3(T)}{3\hbar^2 \delta^3}}{n + \frac{1}{2} + \sqrt{\left(l + \frac{1}{2}\right)^2 + \frac{\mu A_2 m_D^2(T)}{\hbar^2 \delta^3}\left(1 - \frac{m_D(T)}{\delta}\right)}}\right]^2 \tag{31}$$

2. When we set $A_1 = A_2 = 0$, we obtain the energy eigenvalue for Hulthén potential

$$E_{nl} = A_0\left(\frac{1}{2} - \frac{m_D(T)}{4\delta}\right) - \frac{\hbar^2}{8\mu}\left[\frac{\frac{2\mu A_0}{\hbar^2 m_D(T)} - \frac{A_0 \mu m_D(T)}{2\hbar^2 \delta^2}}{n + \frac{1}{2} + \sqrt{\left(l + \frac{1}{2}\right)^2 - \frac{\mu A_0 m_D(T)}{6\hbar^2 \delta^3}}}\right]^2 \tag{32}$$

3. When we set $A_0$, we obtain the energy eigenvalue for Hellmann potential

$$E_{nl} = A_2 m_D(T)\left(\frac{3m_D(T)}{2\delta} - m_D^2(T) - 1\right) - \frac{\hbar^2}{8\mu}\left[\frac{\frac{2\mu}{\hbar^2}(A_1 - A_2) + \frac{3A_2 \mu m_D^2(T)}{\hbar^2 \delta^2} - \frac{8\mu A_2 m_D^3(T)}{3\hbar^2 \delta^3}}{n + \frac{1}{2} + \sqrt{\left(l + \frac{1}{2}\right)^2 + \frac{\mu A_2 m_D^2(T)}{\hbar^2 \delta^3}\left(1 - \frac{m_D(T)}{\delta}\right)}}\right]^2 \tag{33}$$

4. When we set $A_0 = A_2 = m_D(T) = 0$, we obtain the energy eigenvalues for Coulomb potential

$$E_{nl} = -\frac{\mu A_1^2}{2\hbar^2 (n + l + 1)^2} \tag{34}$$

The result of Eq.(34) is very consistent with the result obtained in Eq.(36) of Ref.[33]

To determine the wavefunction, we substitute Eqs.(21) and (24) into Eq.(A4) and obtain

$$\frac{d\phi}{\phi} = \left(\frac{\varepsilon}{x^2\sqrt{\varepsilon}} - \frac{\alpha}{2x\sqrt{\varepsilon}}\right)dx \tag{35}$$

Integrating Eq.(35), we obtain

$$\phi(x) = x^{-\frac{\alpha}{2\sqrt{\varepsilon}}} e^{-\frac{\varepsilon}{x\sqrt{\varepsilon}}} \tag{36}$$

By substituting Eqs.(21) and (24) into Eq.(A6) and integrating, thereafter simplify we obtain

$$\rho(x) = x^{-\frac{\alpha}{\sqrt{\varepsilon}}} e^{-\frac{2\varepsilon}{x\sqrt{\varepsilon}}} \tag{37}$$

Substituting Eqs.(21) and (37) into Eq.(A5) we have

$$y_n(x) = B_n e^{\frac{2\varepsilon}{x\sqrt{\varepsilon}}} x^{\frac{\alpha}{\sqrt{\varepsilon}}} \frac{d^n}{dx^n}\left[e^{-\frac{2\varepsilon}{x\sqrt{\varepsilon}}} x^{2n-\frac{\alpha}{\sqrt{\varepsilon}}}\right] \tag{38}$$

The Rodrigues' formula of the associated Laguerre polynomials is

$$L_n^{\frac{\alpha}{\sqrt{\varepsilon}}}\left(\frac{2\varepsilon}{x\sqrt{\varepsilon}}\right) = \frac{1}{n!} e^{\frac{2\varepsilon}{x\sqrt{\varepsilon}}} x^{\frac{\alpha}{\sqrt{\varepsilon}}} \frac{d^n}{dx^n}\left(e^{-\frac{2\varepsilon}{x\sqrt{x}}} x^{2n-\frac{\alpha}{\sqrt{\varepsilon}}}\right) \tag{39}$$

where

$$\frac{1}{n!} = B_n \tag{40}$$

Hence,

$$y_n(x) \equiv L_n^{\frac{\alpha}{\sqrt{\varepsilon}}}\left(\frac{2\varepsilon}{x\sqrt{\varepsilon}}\right) \tag{41}$$

Substituting Eqs.(36) and (41) into Eq.(A2) we obtain the wavefunction of Eq.(10) in terms of Laguerre polynomial as

$$\psi(x) = B_{nl} x^{-\frac{\alpha}{2\sqrt{\varepsilon}}} e^{-\frac{\varepsilon}{x\sqrt{\varepsilon}}} L_n^{\frac{\alpha}{\sqrt{\varepsilon}}}\left(\frac{2\varepsilon}{x\sqrt{\varepsilon}}\right) \tag{42}$$

where $N_{nl}$ is normalization constant, which can be obtain from

$$\int_0^\infty |B_{nl}(r)|^2 dr = 1 \tag{43}$$

### 3. Results

We calculate mass spectra of the heavy quarkonium system such as charmonium and bottomonium that have the quark and antiquark flavor, and apply the following relation [36,37]

$$M = 2m + E_{nl} \tag{44}$$

where m is quarkonium bare mass and $E_{nl}$ is energy eigenvalues. By substituting Eq.(30) into Eq.(44) we obtain the mass spectra for Hulthén plus Hellmann potential as

$$M = 2m + A_0\left(\frac{1}{2} - \frac{m_D(T)}{4\delta}\right) + A_2 m_D(T)\left(\frac{3m_D(T)}{2\delta} - m_D^2(T) - 1\right)$$

$$-\frac{\hbar^2}{8\mu}\left[\frac{\frac{2\mu}{\hbar^2}\left(A_2 - A_1 + \frac{A_0}{m_D(T)}\right) + \frac{\mu m_D(T)}{\hbar^2 \delta^2}\left(3A_2 m_D(T) - \frac{A_0}{2}\right) - \frac{8\mu A_2 m_D^3(T)}{3\hbar^2 \delta^3}}{n + \frac{1}{2} + \sqrt{\left(l + \frac{1}{2}\right)^2 + \frac{\mu A_2 m_D^2(T)}{\hbar^2 \delta^3}\left(1 - \frac{m_D(T)}{\delta}\right) - \frac{\mu A_0 m_D(T)}{6\hbar^2 \delta^3}}}\right]^2 \quad (45)$$

Table 1
Mass spectra of charmoniumin (GeV) for Hulthén plus Hellmann potential ,($m_c$ =1.209 GeV, $\mu$ = 0.6045 GeV, $A_0$ = -1.693 GeV, $A_1$ = 20.654 GeV, $A_2$ = 0.018 GeV, $\delta$ = 0.2 GeV, $m_D(T)$ = 1.52 GeV, $\hbar$ = 1)

| State | Present work | [35] | [24] | Experiment[38,39] |
|---|---|---|---|---|
| 1S | 3.096 | 3.096 | 3.096 | 3.096 |
| 2S | 3.686 | 3.686 | 3.672 | 3.686 |
| 1P | 3.521 | 3.255 | 3.521 | 3.525 |
| 2P | 3.772 | 3.779 | 3.951 | 3.773 |
| 3S | 4.040 | 4.040 | 4.085 | 4.040 |
| 4S | 4.262 | 4.269 | 4.433 | 4.263 |
| 1D | 3.768 | 3.504 | 3.800 | 3.770 |
| 2D | 4.146 | - | - | 4.159 |
| 1F | 3.962 | - | - | - |

Table 2
Mass spectra of bottomonium in (GeV) for Hulthén plus Hellmann potential,($m_b$ =4.823 GeV, $\mu$ = 2.4115 GeV, $A_0$ = -1.591 GeV, $A_1$ = 9.649 GeV, $A_2$ = 0.028 GeV, $\delta$ = 0.25 GeV, $m_D(T)$ = 1.52 GeV, $\hbar$ = 1)

| State | Present work | [35] | [24] | Experiment[38,39] |
|---|---|---|---|---|
| 1S | 9.460 | 9.460 | 9.462 | 9.460 |
| 2S | 10.023 | 10.023 | 10.027 | 10.023 |
| 1P | 9.861 | 9.619 | 9.9630 | 9.899 |
| 2P | 10.238 | 10.114 | 10.299 | 10.260 |
| 3S | 10.355 | 10.355 | 10.361 | 10.355 |
| 4S | 10.579 | 10.567 | 10.624 | 10.580 |
| 1D | 10.143 | 9.864 | 10.209 | 10.164 |

| | | | | | |
|---|---|---|---|---|---|
| 2D | 10.306 | - | - | - | |
| 1F | 10.209 | - | - | - | |

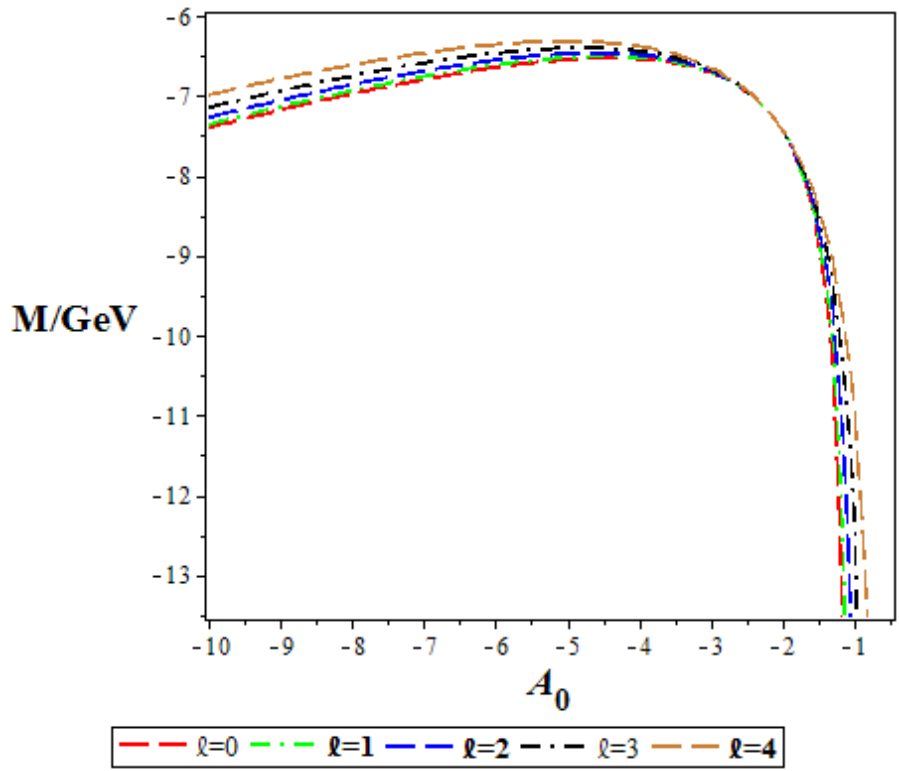

FIGURE 1. Mass spectra variation with potential parameter $A_0$ for different quantum numbers

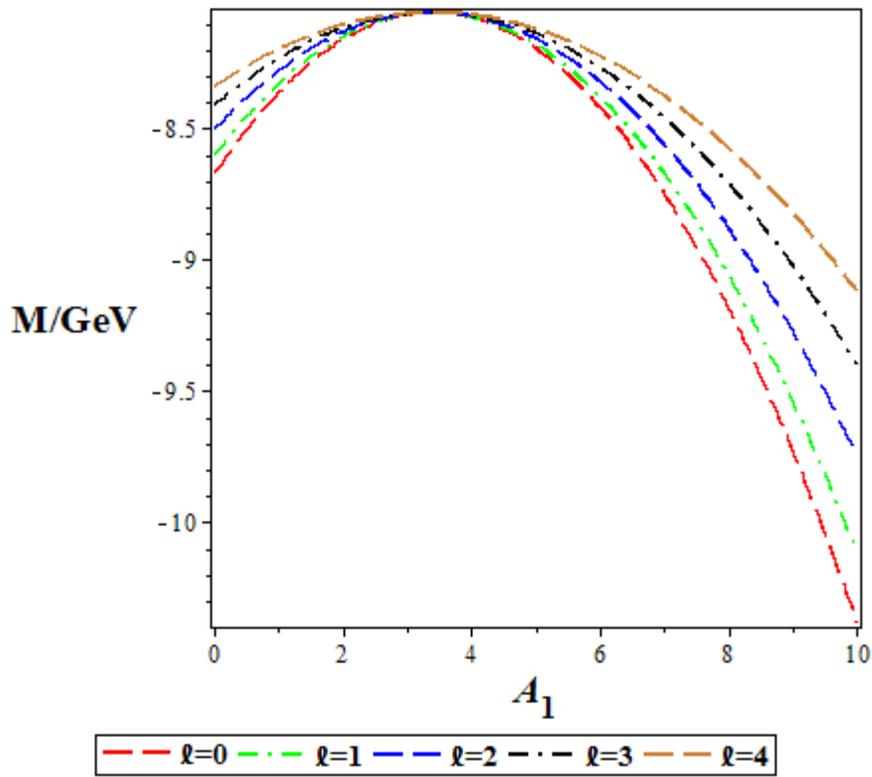

FIGURE 2. Mass spectra variation with potential parameter $A_1$ for different quantum numbers

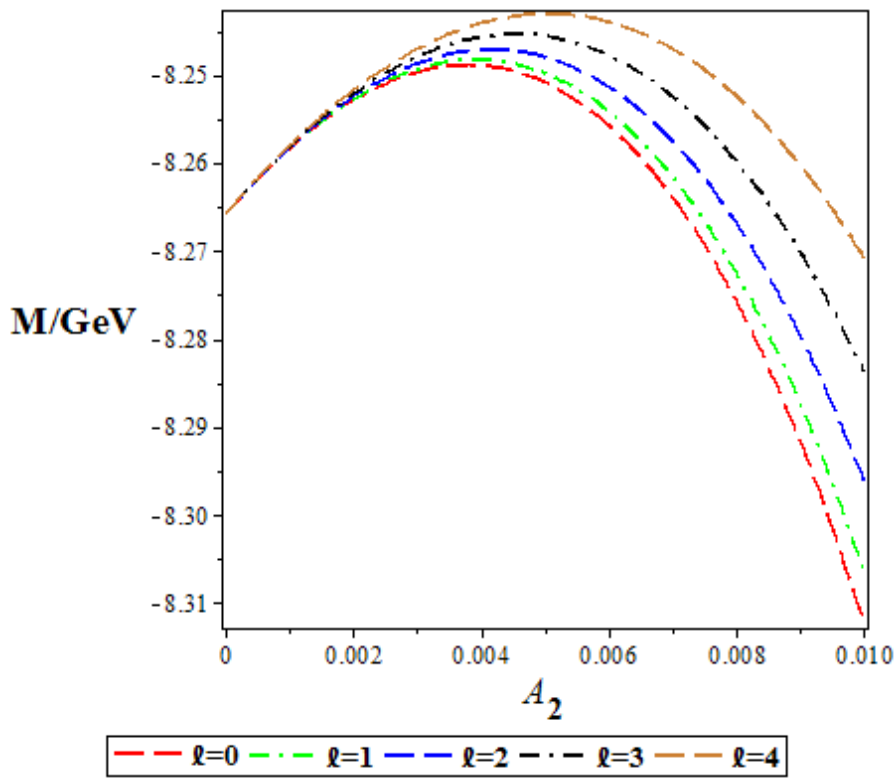

FIGURE 3. Mass spectra variation with potential parameter ($A_2$) for different quantum numbers

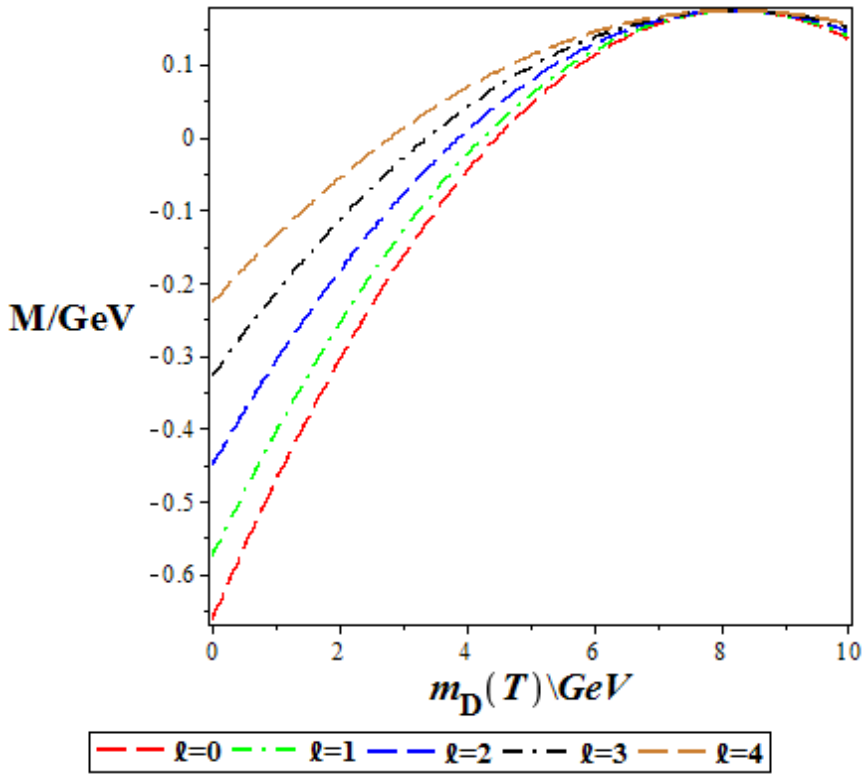

FIGURE 4. Mass spectra variation with the Debye mass $m_D(T)$ for different quantum numbers

3.1 Discussion of results

We calculate mass spectra of charmonium and bottomonium for states from 1S to 1F, by using Eq. (45). The free parameters of Eq. (45) are fitted with experimental data by solving two algebraic equations. Experimental data are obtained from [38,39]. For bottomonium $b\bar{b}$ and charmonium $c\bar{c}$ systems we adopt the numerical values of these masses as $m_b = 4.823\,GeV$ and $m_c = 1.209\,GeV$, respectively [40]. Then, the corresponding reduced mass are $\mu_b = 2.4115\,GeV$ and $\mu_c = 0.6045\,GeV$. The Debye mass $m_D(T)$ is taken as $1.52\,GeV$ by fitted with experimental data. We note that calculation of mass spectra of charmonium and bottomonium are in a good agreement with experimental data as well as the work of other researchers as presented in Tables 1 and 2. It's important to note that the values obtained are improved in comparison with works of other researcher like; Ref.[35] as shown in Tables 1 and 2 in which the author investigated the N- radial SE analytically when the Cornell potential was extended to finite temperature.

We also plotted the mass spectra energy as a function of potential parameters and Debye mass .In Figs. 1 and 2, the mass spectra energies increases to a peak and later decreases as potential parameters $A_0$ and $A_1$ increases, respectively. In Fig.3 the mass spectra converges at the beginning, but spreads out and decrease monotonically with the increase in potential parameter $A_2$. Figure 4 show the increase in mass spectra as the Debye mass increases, for various angular quantum numbers.

4. Conclusion

In this study, we adopted Hulthén plus Hellmann potential models for quark-antiquark interaction. The potential was made to be temperature dependent by replacing the screening parameter with Debye mass $(m_D(T))$ which vanishes at

$T \to 0$. The Schrödinger equation is analytically solved using Nikiforov-Uvarov method. We obtained approximate solutions of the eigenvalues and eigenfunction in terms of Laguerre polynomials. We applied the present results to calculate heavy-meson masses such as charmonium $c\bar{c}$, and bottomonium $b\bar{b}$ for states 1S to 1F which are in good agreement with experimental data and the work of others. Four special cases were considered when some of the potential parameters were set to zero, resulting into Hellmann potential, Yukawa potential, Coulomb potential, and Hulthén potential, respectively. Different plots of mass spectra versus different potential parameters and Debye mass were analyzed and discussed.

**References**

1. H. Mutuk, Mass Spectra and Decay constants of Heavy-light Mesons: A case study of QCD sum Rules and Quark model, *Adv. in High Energy Phys.* **20** (2018) 8095653.

2. A. Mocsy, Potential models for quarkonia. *The Euro. Phys. J.* **61** (2009)710.

3. A. K. Rai, and D. P. Rathaud, The mass spectra and decay properties of dimesonic states, using the Hellmann potential, *Eur. Phys. J C* **75** (2015) 9. doi:10.1140/epjc/s10052-015-3695-z

4. A. N. Ikot, U. S. Okorie, G. Osobonye, P.O. Amadi, C. O. Edet, M. J. Sithole, G. J. Rampho, and R. Sever, Superstatistics of Schrodinger equation with pseudo-harmonic potential in external magnetic and Aharanov-Bohm fields, *Heliyon* **6** (2020) e03738, doi: 10.1016/j.heliyon.2020.e03738

5. S. M. Ikhdair, Relativistic bound states of spinless particle by the Cornell potential model in in external fields. *Adv. in High Energy Phys.* (2013) 491648

6. A. Mira-Cristiana,Yukawa:The man and the potential. *Didactica Mathematica*, **33** (2015) 9.

7. E. Omuge, O. E. Osafile, and M. C. Onyeajh, Mass spectrum of mesons via WKB Approximation method. *Adv. in High Ener. Phys.***10** (2020) 1143.

8. H. Mansour, and A. Gamal, Bound state of Heavy Quarks using a General polynomial potential. *Adv. in High Ener. Phys.* **65** (2018) 1234.

9. L. Hulthén, Uber die eigenlosunger der schrodinger - gleichung des deuterons, *Ark. Mat. Astron. Fys. A* **28** (1942) 5

10. T. Xu, Z. Q. Cao, Y. C. Ou, Q. S. Shen, and G. L. Zhu, Critical radius and dipole polarizability for a confined system. *Chine. Phys*. **15** (2006) 1172

11. B. Durand, and L. Durand, Duality for heavy-quark systems. *Phys. Rev. D*, **23** (1981)1092.

12. C. S. Jia, J. Y. Wang, S. He, and L. T. Sun, Shape invariance and the supersymmetry WKB approximation for a diatomic molecule potential. *J. Phys. A*, **33** (2000) 6993.

13. H. Hellmann, A New Approximation Method in the Problem of Many Electrons, *J. Chem Phys.* **3** (1935) 61. https://doi.org/10.1063/1.1749559

14. B. J. Falaye ,K.J. Oyewumi, T. T. Ibrahim, M. A. Punyasena, and C. A. Onate, Bound solutions of the Hellmann potential. *Canad. J Phys.* **91** (2013) 98.

15. C. A. Onate, O. Ebomwonyi, K. O. Dopamu, J. O. Okoro, and M. O. Oluwayemi, Eigen solutions of the D-dimensional Schrödinger equation with inverse trigonometry scarf potential and Coulomb potential, *Chin. J. Phys*, (2018), doi:10.1016/j.cjph.2018.03.013

16. E. S. William, J. A. Obu, I. O. Akpan, E. A. Thompson and E. P. Inyang, Analytical Investigation of the Single-particle energy spectrum in Magic Nuclei of $^{56}$Ni and $^{116}$Sn. *European Journal of Applied Physics* **2** (2020) 28 http://dx.doi.org/10.24018/ejphysics.2020.2.6.28

17. S. M. Ikhdair, An improved approximation scheme for the centrifugal term and the Hulthén potential. *The Eur. Phys. J. A,* **39** (2009) 3, 307–314. doi:10.1140/epja/i2008-10715-2

18. E. P. Inyang, E. S. William and J. A. Obu, Eigensolutions of the N-dimensional Schrödinger equation interacting with Varshni-Hulthen potential model. *Rev. Mex. Fis*. arXiv:2012.13826,(2020)

19. J. E. Ntibi, E. P. Inyang, E. P. Inyang, and E. S. William, Relativistic Treatment of D-Dimensional Klien-Gordon equation with Yukawa potential . *Intl J. of Innov Sci, Engr. Tech* **11** (2020) .

20. M. Hamzavi, K. E. Thylwe, and A. A. Rajabi, Approximate Bound States Solution of the Hellmann Potential, *Commun Theor Phys,* **60** (2013) *8.* doi:10.1088/0253-6102/60/1/01

21. E.P. Inyang , E.P. Inyang, J. E. Ntibi, E. E. Ibekwe, and E. S. William, Approximate solutions of D-dimensional Klein-Gordon equation with Yukawa potential via Nikiforov-Uvarov method**.** *Ind. J. Phys*. **20** (2020).https://doi.org/10.1007/s12648-020-01933-x

22. A.Vega and J.Flores, Heavy quarkonium properties from Cornell potential using variational method and supersymmetric quantum mechanics, *Pramana J.Phys*. **87** (2016).

23. M. Abu-Shady, T. A. Abdel-Karim, and Y. Ezz-Alarab, Masses and thermodynamic properties of heavy mesons in the non-relativistic quark model using the Nikiforov-Uvarov method. J. of Egypt. Math. Soci. **27** (2019)145.

24. H. Ciftci, and H. F. Kisoglu, Non-relativistic Arbitary $l$ -states of Quarkonium through Asymptotic interation method. *Adv. in High Ener. Phys*. **45** (2018) 49705.https://doi.org/10.1155/2018/4549705

25. M. Abu-Shady, Analytic solution of Dirac Equation for extended Cornell Potential using the Nikiforov-Uvarov method. *Bos. J. mod. Phys*. **55** (2015) 789.

26. A. Al-Jamel, and H. Widyan, Heavy quarkonium mass spectra in a Coulomb field plus Quadratic potential using Nikiforov-Uvarov method. *Appl. Phys. Research* **4** (2012).

27. E. E. Ibekwe , T. N. Alalibo, S. O. Uduakobong , A. N. Ikot, and N.Y.Abdullah, Bound state solution of radial Schrödinger equation for the quark-antiquark interaction potential. *Iran J. of Sci. Tech*. **20** (2020) 00913.

28. A. Al-Oun, A. Al-Jamel, and H. Widyan, Various properties of Heavy Quakonium from Flavor-independent Coulomb plus Quadratic potential. *Jord. J. Phys*., **40** (2015) 453.

29. A. Chouikh,T. Said, and M. Bennai, Alternative Approach for Quantum Computation in a cavity QED. *Quant.Phys.Lett*. **6** (2017) 65.

30. R. Kumar, and F. Chand, Solutions to the N-dimensional radial Schrodinger equation for the potential $ar^2 + br - c/r$. *Pramana. J. of Phys*. **34** (2014) 10.

31. C. A. Onate, J. O. Ojonubah, A. Adeoti, E. J. Eweh, and M. Ugboja, Approximate Eigen Solutions of D.K.P. and Klein-Gordon Equations with Hellmann Potential, *Afr. Rev. Phys*, **497** (2014) 8.

32. E. S. William, E. P.Inyang, and E. A. Thompson, Arbitrary $\ell$ -solutions of the Schrödinger equation interacting with Hulthén-Hellmann potential model. *Rev. Mex. Fis*. **66** (2020) 730. https://doi.org/10.31349/RevMexFis.66.730.

33. C. O. Edet, U. S. Okorie, A. T. Ngiangia, and A. N. Ikot, Bound state solutions of the Schrödinger equation for the modified Kratzer plus screened Coulomb potential. *Ind. J. Phys*. **94** (2020)423.

34. E. P. Inyang, J. E. Ntibi, E. P. Inyang, E. S. William and C. C. Ekechukwu, Any L- state solutions of the Schrödinger equation interacting with class of Yukawa - Eckert potentials. *Intl J. Innov Sci, Engr. Tech* **7** (2020)

35. M. Abu-Shady, N-dimensional Schrödinger equation at finite temperature using the Nikiforov-Uvarov method. *J. Egypt. Math. Soci.* **23** (2016) 4.

36. E. P. Inyang, E. P. Inyang, E. S. William, E. E. Ibekwe and I. O. Akpan, Analytical Investigation of Meson Spectrum via Exact Quantization Rule Approach.arXiv:2012.10639,(2020)

37. E. P.Inyang, E. P. Inyang, I. O. Akpan and E. S. William, Analytical solutions of the Schrödinger equation with class of Yukawa potential for a Quarkonium system via series expansion method. European Journal of Applied Physics **2** (2020) 26 http://dx.doi.org/10.24018/ejphysics.2020.2.6.26

38. M. Tanabashi, C. D. Carone,T.G.Trippe, and C.G. Wohl, Particle Data Group. *Phys. Rev. D,* **98** (2018)546.

39. R. Olive, D. E. Groom, and T. G. Trippe, Particle Data Group. *Chin. Phys. C,***38** (2014) 54.

40. R. M. Barnett, C. D. Carone, D. E. Groom, T. G. Trippe, and C.G.Wohl, Particle Data Group. *Phys. Rev. D*, **92** (2012) 654.

41. A. F. Nikiforov, V. B. Uvarov, *Special Functions of Mathematical Physics*, (Birkhauser, Bassel, 1988)

42. C. M .Ekpo, E. P. Inyang, P. O. Okoi, et al., New Generalized Morse-Like Potential for studying the Atomic Interaction in Diatomic Molecules. http://arXiv:2012.02581(2020)

**APPENDIX A: Review of Nikiforov-Uvarov (NU) method**

The NU method was proposed by Nikiforov and Uvarov [41,42] to transform Schrödinger-like equations into a second-order differential equation via a coordinate transformation $x = x(r)$, of the form

$$\psi''(x) + \frac{\tilde{\tau}(x)}{\sigma(x)}\psi'(s) + \frac{\tilde{\sigma}(x)}{\sigma^2(x)}\psi(x) = 0 \tag{A1}$$

where $\tilde{\sigma}(x)$, and $\sigma(x)$ are polynomials, at most second degree and $\tilde{\tau}(x)$ is a first-degree polynomial. The exact solution of Eq.(A1) can be obtain by using the transformation.

$$\psi(x) = \phi(x) y(x) \tag{A2}$$

This transformation reduces Eq.(A1) into a hypergeometric-type equation of the form

$$\sigma(x) y''(x) + \tau(x) y'(x) + \lambda y(x) = 0 \tag{A3}$$

The function $\phi(x)$ can be defined as the logarithm derivative

$$\frac{\phi'(x)}{\phi(x)} = \frac{\pi(x)}{\sigma(x)} \tag{A4}$$

With $\pi(x)$ being at most a first-degree polynomial. The second part of $\psi(x)$ being $y(x)$ in Eq.(A2) is the hypergeometric function with its polynomial solution given by Rodrigues relation as

$$y(x) = \frac{B_{nl}}{\rho(x)} \frac{d^n}{dx^n}\left[\sigma^n(x)\rho(x)\right] \tag{A5}$$

Where $B_{nl}$ is the normalization constant and $\rho(x)$ the weight function which satisfies the condition below;

$$(\sigma(x)\rho(x))' = \tau(x)\rho(x) \tag{A6}$$

where also

$$\tau(x) = \tilde{\tau}(x) + 2\pi(x) \tag{A7}$$

For bound solutions, it is required that

$$\tau'(x) < 0 \tag{A8}$$

The eigenfunctions and eigenvalues can be obtained using the definition of the following function $\pi(x)$ and parameter $\lambda$, respectively:

$$\pi(x) = \frac{\sigma'(x) - \tilde{\tau}(x)}{2} \pm \sqrt{\left(\frac{\sigma'(x) - \tilde{\tau}(x)}{2}\right)^2 - \tilde{\sigma}(x) + k\sigma(x)} \tag{A9}$$

and

$$\lambda = k_- + \pi'_-(x) \tag{A10}$$

The value of $k$ can be obtained by setting the discriminant in the square root in Eq. (A9) equal to zero. As such, the new eigenvalues equation can be given as

$$\lambda + n\tau'(x) + \frac{n(n-1)}{2}\sigma''(x) = 0, (n = 0, 1, 2, ...) \tag{A11}$$